\documentclass[prd,aps]{revtex4}

\usepackage{epsfig}

\newcommand{\beq}{\begin{equation}}
\newcommand{\eeq}{\end{equation}}
\newcommand{\be}{\begin{eqnarray}}
\newcommand{\ee}{\end{eqnarray}}

\begin{document}



\title{
Lepton flavour violation in a nonuniversal gauge interaction model 
}

\author{ Kang Young Lee }
\email{kylee14214@gmail.com}
\vskip 0.5cm

\affiliation{
Division of Quantum Phases \& Devices, 
School of Physics, Konkuk University, Seoul 143-701, Korea
}

\date{\today}

\begin{abstract}

The flavour-changing neutral currents (FCNC) are derived at tree level 
if the electroweak gauge group depends on the fermion family,
which are absent in the standard model.
We study the lepton flavour violation (LFV) through the FCNC interactions
in a nonuniversal gauge interaction model 
where the third generation fermions are subjected to 
the separate SU(2)$_L$ gauge group.

\end{abstract}

\pacs{PACS numbers: 12.60.Cn,13.35.-r}

\maketitle






Search for signatures of new physics beyond the standard model (SM) 
is the most important task of the particle physics at present.
The lepton family number is exactly preserved in the SM
and no lepton flavour violation (LFV) has been observed 
at experiments so far.
The experimental bounds on the LFV in $Z$ decay
are given by 
\be
{\rm Br}(Z \to e \mu) < 1.7 \times 10^{-6},
\nonumber \\
{\rm Br}(Z \to e \tau) < 9.8 \times 10^{-6},
\nonumber \\
{\rm Br}(Z \to \mu \tau) < 1.2 \times 10^{-5},
\ee
at 95\% C.L. \cite{pdg}
and those in $\tau$ decays given by
\be
{\rm Br}(\tau^- \to e^- e^+ e^-) < 3.6 \times 10^{-8},
\nonumber \\
{\rm Br}(\tau^- \to e^- \mu^+ \mu^-) < 3.7 \times 10^{-8},
\nonumber \\
{\rm Br}(\tau^- \to e^+ \mu^- \mu^-) < 2.3 \times 10^{-8},
\nonumber \\
{\rm Br}(\tau^- \to \mu^- e^+ e^-) < 2.7 \times 10^{-8},
\nonumber \\
{\rm Br}(\tau^- \to \mu^+ e^- e^-) < 2.0 \times 10^{-8},
\nonumber \\
{\rm Br}(\tau^- \to \mu^- \mu^+ \mu^-) < 3.2 \times 10^{-8},
\ee
and those in muon decays 
\be
{\rm Br}(\mu^- \to e^- e^+ e^-) < 1.0 \times 10^{-12},
\ee
at 90\% C.L.
\cite{pdg}.
Although existence of the neutrino oscillations implies that 
the lepton family numbers are no more preserved,
the LFV through neutrino oscillation are 
too suppressed to be observed due to the smallness of neutrino mass.
Therefore observation of the LFV would be an unambiguous evidence 
for deviations from the SM.

We consider the model with a separate SU(2) group
acting only on the third generation fermions, 
in which the LFV generically arises.
Since different SU(2) gauge group is assigned on the third generation,
the corresponding gauge coupling constant is different
from that of the first and second generations in general,
which indicates that the electroweak gauge interaction
is nonuniversal in this model.
In the SM, the neutral current interactions are simultaneously
diagonalized with mass matrices of quarks and leptons 
by a biunitary transformation with two independent unitary matrices
corresponding to left-handed and right-handed quarks and leptons,
which results in the absence of the flavour changing neutral current (FCNC)
at tree level.
However, if the SU(2) gauge couplings are nonuniversal,
the neutral current interactions cannot be diagonalized 
simultaneously with mass matrices
and FCNCs exist in general.
The FCNC of the lepton sector means violation 
of the lepton flavours.

The phenomenology of our model has been studied 
using the $Z$-pole data and the low-energy neutral current data 
\cite{topflavour1,topflavour2}.
Phenomenological implications of the FCNC interactions of this model
have been considered in Ref. \cite{topflavour2,fcnc}.
By the way the nonuniversality of the gauge couplings in this model 
also leads to the violation of the unitarity 
of the Cabibbo-Kobayashi-Maskawa (CKM) matrix \cite{kylee3,kylee4}.
The symmetry breaking mechanism of this model has been studied 
in detail in Ref. \cite{chiang}.
This gauge group can arise as the theory
at an intermediate scale in the path of gauge
symmetry breaking of noncommuting extended technicolor
(ETC) models \cite{ETC} 
and the grand unified theory based on SU(3)$^3$ or SU(15).

The LFV in this model has been discussed in Ref \cite{topflavour2}.
However they assumed the case that one mixing angle
between two flavours is dominant and just analyzed corresponding channels.
In this paper, we present a general description of the LFV
in SU(2)$_l \times$ SU(2)$_h \times$ U(1) model and 
perform the complete analysis with the recent experimental data.
We show that the flavour conserving $Z \to l^-_i l^+_i$ decays
play crucial roles for constraining the model
as well as the strong limits on the LFV processes.


We consider the electroweak $SU(2)_l \times SU(2)_h \times U(1)_Y $ model,
in which the gauge group $SU(2)_l$ acts on 
the first and the second generations 
and $SU(2)_h$ on the third generation only.
The quantum numbers of left-handed quarks and leptons
are assigned as (2,1,Y) for the first and second generations and 
(1,2,Y) for the third generation
under $SU(2)_l \times SU(2)_h \times U(1)_Y $.

The spontaneous symmetry breaking mechanism is achieved 
such that the gauge group $ SU(2)_l \times SU(2)_h \times U(1) $
breaks down into the $ SU(2)_{l+h} \times U(1)_Y $
with the vacuum expectation values (VEV)
of the (2,2,0) bidoublet scalar field $\Sigma$,
$\langle \Sigma \rangle = {\rm diag}(u,u)$
and sequentially breaks down into $U(1)_{em}$
with the VEV of the scalar doublet $\Phi$, 
$\langle \Phi \rangle = (0, v/\sqrt{2})$.
Since the $ SU(2)_l \times SU(2)_h $ breaking scale $u$ is
higher than the electroweak scale $v$, 
we introduce a small parameter $\lambda \equiv v^2/u^2 \ll 1 $.
The masses of the heavy gauge bosons are degenerate in this model,
$m_{_{W'^{\pm}}}^2 = m_{_{Z'}}^2
= m_0^2/\lambda \sin^2 \phi \cos^2 \phi$ ,
while the ordinary $W$ boson masses given by
$ m_{_{W^{\pm}}}^2
= m_0^2 ( 1- \lambda \sin^4 \phi )
= m_{_{Z}}^2 \cos^2 \theta$
where $m_0 = ev/(2 \sin \theta)$.

We parameterize the gauge couplings by
$g_l = e/\sin \theta \cos \phi$,
$g_h = e/\sin \theta \sin \phi$, and
$g^{\prime} = e/\cos \theta$
in terms of the electromagnetic coupling $e$, 
the weak mixing angle $\theta $ and the new mixing angle $\phi$ 
between $SU(2)_l$ and $SU(2)_h$.
The perturbativity $g_{(l,h)}^2/4 \pi < 1$ is assumed,
which results in the constraint $ 0.03 < \sin^2 \phi < 0.96 $.

We focus on the neutral current interactions in this paper.
It is advantageous to separate the nonuniversal part
from the universal part in the Lagrangian
such that
${\cal L}^{\rm NC} =  {\cal L}^{\rm NC}_{I} + {\cal L}^{\rm NC}_{3}$,
where $ {\cal L}^{\rm NC}_{I} $ denotes the universal gauge interactions
and $ {\cal L}^{\rm NC}_{3} $ the nonuniversal terms.
The universal part is given by
\be
{\cal L}^{\rm NC}_{I} = 
\bar{f}_L \gamma_\mu (G_L Z^\mu + G'_L Z'^\mu) f_L
+ \bar{f}_R \gamma_\mu (G_R Z^\mu + G'_R Z'^\mu) f_R,
\ee 
where $f = (e,\mu,\tau)^T$ for the lepton sector and 
\be
G_L &=& -\frac{e}{\cos \theta \sin \theta}
         \left( T_3 - Q \sin^2 \theta - \lambda T_3 \sin^4 \phi 
         \right) I,
\nonumber \\
G'_L &=& \frac{e}{\sin \theta} 
           \left( T_3 \tan \phi 
                   + \lambda \frac{\sin^3 \phi \cos \phi}{\cos^2 \theta} 
                              (T_3 - Q \sin^2 \theta) 
            \right) I, 
\nonumber \\
G_R &=& \frac{e}{\cos \theta \sin \theta} Q \sin^2 \theta \cdot I,
\nonumber \\
G'_R &=& -\frac{e}{\sin \theta} \lambda Q \tan^2 \theta 
                                 \sin^3 \phi \cos \phi \cdot I,
\ee
with the 3$\times$3 identity matrix $I$
and the electric charge defined by $ Q = T_{3l} + T_{3h} + Y/2 $.
The nonuniversal part consists of a nonzero (3,3) elements
which is given by
${\cal L}^{\rm NC}_{3} = 
\bar{f}_L \gamma_\mu (Y_L Z^\mu + Y'_L Z'^\mu) f_L, $
where 
\be
Y_L &=& -\frac{e}{\cos \theta \sin \theta} \lambda T_3 \sin^2 \phi \cdot M,
\nonumber \\
Y'_L &=& -\frac{e}{\sin \theta} 
            \frac{T_3}{ \sin \phi \cos \phi} M, 
\ee
with the $3\times3$ matrix $M$ defined by $M_{ij} = \delta_{3i}\delta_{3j}$.

The mass eigenstates of charged leptons $f^0_{L,R}$ 
by the relation $f_{L,R} = V_{L,R}^\dagger f^0_{L,R}$ 
with the unitary matrix $V_{L,R}$.
The left-handed neutral current interactions are derived to be
\be
{\cal L}^{NC}_L = 
G_L \bar{f}^0_L \gamma_\mu
        \left( I + \epsilon V^\dagger M V \right)
            f^0_L Z^\mu
+ G'_L \bar{f}^0_L \gamma_\mu
        \left( I + \epsilon' V^\dagger M V \right)
            f^0_L Z'^\mu,
\ee
in terms of mass eigenstates,
where $\epsilon$ and $\epsilon'$ are given by
\be
\epsilon &=& \frac{Y_L}{G_L} 
= \frac{\lambda \sin^2 \phi}{1-2 \sin^2 \theta} + {\cal O}(\lambda^2),
\nonumber \\
\epsilon' &=& \frac{Y'_L}{G'_L} 
= -\frac{1}{\sin^2 \phi} + {\cal O}(\lambda),
\ee
while the right-handed neutral current interactions 
are kept to be diagonal.
Thus $V$ is just $V_L$ here.
Since $(V^\dagger M V)_{ij} = V^*_{3i}V_{3j} \ne \delta_{ij}$
in ${\cal L}^{\rm NC}_{L}$, 
the FCNC interaction terms generically arise
at tree level in this model.
Note that the matrix elements, $|V_{31}|$, $|V_{32}|$, $|V_{33}|$, 
are observables in this model.
We will drop the superscript $0$ for mass eigenstates hereafter.
The FCNC parameter $\epsilon$ is of order $\lambda$
and suppress the FCNC interactions.
We note that $\epsilon'$ does not contain a suppression parameter but
the term involving $\epsilon'$ is suppressed
by the heavy mass $m_{Z'}$.

Our model has been examined by various experimental data so far.
The electroweak precision test with the data at the $Z$-pole
has provided strong constraints on new physics models.
We have used the nonstandard electroweak precision test 
with $\epsilon_i$ variables to examine the our model,
which is introduced by Altarelli \cite{sm2,altarelli}
and extended with a new parameter $\epsilon'_b$
to measure the new effects in $Z b \bar{b}$ coupling \cite{kylee5}.
The new physics effects at tree level are assumed to be 
compatible to the electroweak corrections in the SM such that
$\epsilon_i = \epsilon_i^{\rm SM} + \epsilon_i^{\rm new}$.
The SM predictions $ \epsilon_i^{\rm SM} $
are calculated by ZFITTER \cite{ZFITTER}
and $\epsilon_i^{\rm new}$ can be read out
from the $Z f \bar{f}$ couplings in Eq. (7).
With the LEP and SLC data at $Z$-pole \cite{LEPEWWG},
we obtain the constraints on the model parameters
$(\lambda, \sin^2 \phi)$ in Ref. \cite{kylee3,kylee4}.

The precisely measured data 
for low-energy neutral current interaction processes 
such as $\nu e\rightarrow \nu e$,  $\nu N$ scattering,
and the polarized electron-hadron scattering $e_{L,R} N \rightarrow eX$
can examine our model 
through corrections to the ordinary $Z$ boson couplings 
and the extra $Z'$ boson contribution.
The coefficients of the effective four-fermion Hamiltonian
are generically shifted by
$C = C_{\rm SM} (1-\lambda \sin^4 \phi)$
in this model. 
The atomic parity violation is described by the same Hamiltonian 
as the $e_{L,R} N \rightarrow eX$ scattering.
The weak charge of an atom defined by 
$Q_W = -2 \Bigl[ C_{1u} (2Z+N) + C_{1d} (Z+2N) \Bigr]$
is also corrected by $ \Delta Q_W \equiv Q_W - Q_W^{\mathrm{SM}}
= \Delta Q_W^{\rm SM} (1 - \lambda \sin^4 \phi)$,
where $Z$ ($N$) is the number of protons (neutrons) in the atom. 
With the recent measurement of the weak charge for the Cs an Tl atoms 
\cite{apv_exp},
and the SM predictions $ Q_W = -73.19 \pm 0.03$ (Cs),
and $ Q_W = -116.81 \pm 0.04$ (Tl),
constraints on the model parameters 
have been obtained in Ref. \cite{kylee3,kylee4}.

The charged current interactions are also affected by
the nonuniversality of the SU(2) couplings 
as well as the neutral currents. 
For the quark sector, 
the charged current couplings are measured
by the CKM matrix elements.
The `observed' CKM matrix is defined by the couplings 
in the low-energy four fermion effective Hamiltonian
describing the semileptonic quark decay.
In this model, 
we extract `observed' $V_{CKM}$ as
\be
V_{CKM} = V_{CKM}^0 + \epsilon^c {V_U}^\dagger M V_D
              + \left( \frac{G'^c_L}{G^c_L} \right)^2
                \frac{m_W^2}{m_{W'}^2}
     \left( V_{CKM}^0 + \epsilon'^c {V_U}^\dagger M V_D \right)
\ee
where $V_U$ and $V_D$ are unitary matrices 
which diagonalize up- and down-type quarks, 
$V_{CKM}^0 \equiv {V_U}^\dagger V_D$ is the CKM matrix in the SM,
and $M = \delta_{3i} \delta_{3j}$ are given previously.
Using the correction terms 
$\epsilon^c = \lambda \sin^2 \phi + {\cal O}(\lambda^2)$
and $\epsilon'^c = 1/\sin^2 \phi + {\cal O}(\lambda)$,
we have the simple form of $V_{CKM}=V_{CKM}^0 (1 + \lambda \sin^4 \phi)$.
Since the correction term violates the unitarity of the CKM matrix,
test of the CKM unitarity can examine our model.
We define the unitarity violating term $\Delta$ by
the normalization of the first row of the CKM matrix,
$|V_{ud}|^2 + |V_{us}|^2 + |V_{ub}|^2 = 1-\Delta$,
which is the most precise test of the unitarity of the CKM matrix.
We derive $\Delta = 2 \lambda \sin^4 \phi$ in our model 
while $\Delta=0$ exactly in the SM.
Recent measurements of nuclear beta decays, kaon decays and $B$ decays
result in $\Delta = 0.0009 \pm 0.0010$
implying the unitarity of the CKM matrix with high accuracy
\cite{czarnecki}.
We show that the CKM unitarity leads to the stronger
constraint than the electroweak precision test with the LEP and SLC data
and the low-energy neutral current data
on the parameter space ($\sin^2 \phi, m_{Z'}$)
as shown in Fig.1 of Ref. \cite{kylee3}.

\begin{figure}[t]
\begin{center}
\hbox to\textwidth{\hss\epsfig{file=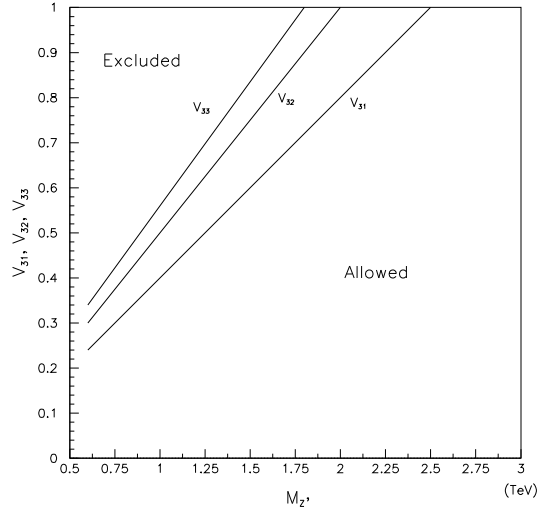,width=8cm}\hss}
\caption{
Bounds on $V_{31}$, $V_{32}$, and $V_{33}$
with respect to $m_{Z'}$ from the $Z \to l^- l^+$ decays.
}
\end{center}
\end{figure}

In the lepton sector, the FCNC interactions given in Eq. (7) 
yield the lepton number violating processes at tree level.
Before studying the LFV processes, 
we notice that the partial decay rates of $Z$ boson into lepton pairs 
are also received the flavour conserving corrections 
from the diagonal elements of $V^\dagger M V$.
In this model, the partial decay rates are shifted as
\be
\Gamma(Z \to l_i^- l_i^+) 
= \Gamma_{\rm SM} \left( 1 + 2 \epsilon |V_{3i}|^2 \right),
\ee
in the leading order of $\lambda$.
The precisely measured data of $\Gamma(Z \to l_i^- l_i^+)$ 
without lepton universality by LEP
is given in Table 7.1 of Ref. \cite{LEPEWWG}.
We obtain constraints on $\epsilon$ and $|V_{3i}|$ from the data.
Figure 1 shows that allowed values of $V_{31}$, $V_{32}$, and $V_{33}$
with respect to $m_{Z'}$ by LEP data varying all possible values of
$\sin^2 \phi$.
We find that conservative bounds on $V_{3i}$ can be achieved
for given value of $m_{Z'}$.

The LFV $Z$ decays are described by
\be
\Gamma(Z \to l^-_i l^+_j) = \Gamma(Z \to l_i^+ l_i^-) 
\cdot \epsilon^2 |V_{3i}|^2 |V_{3j}|^2,
\ee
neglecting lepton masses.
Assuming that $\sin^2 \phi \sim 0.1$, 
we have $\epsilon < 0.02$ which leads to
${\rm Br}(Z \to l^+_i l^-_j) < |V_{3i}|^2 |V_{3j}|^2 \times 10^{-5}$.
Since both $|V_{3i}|$ and $|V_{3j}|$ cannot be 1,
the predictions of ${\rm Br}(Z \to l^+_i l^-_j)$ in our model
is likely to satisfy the present experimental bounds.

\begin{figure}[t]
\begin{center}
\hbox to\textwidth{\hss\epsfig{file=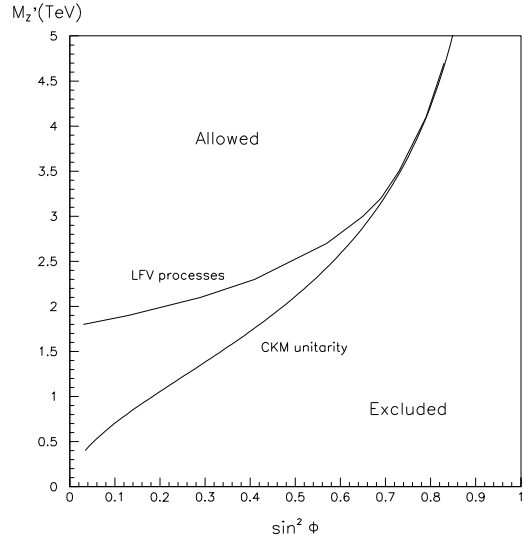,width=8cm}\hss}
\caption{
Bounds on $(\sin^2 \phi, m_{Z'})$ from the CKM unitarity, the LFV bounds
and the partial $Z$ decays into lepton pairs.
}
\end{center}
\end{figure}

The LFV lepton decays occur through the $Z$ and $Z'$ exchanges.
With the FCNC couplings derived in Eq. (7),
the flavour violating $\tau$ decay rates are given by
\be
\Gamma(\tau^- \to \mu^- (e^-) \mu^+ \mu^-) 
&=& \frac{m_\tau^5}{96 \pi^3} \left[ 
\left| G_{4LL}^{23(13)} + {G'}_{4LL}^{23(13)} \right|^2 
+ \left| G_{4LR}^{23(13)} \right|^2 \right],
\nonumber \\ 
&=& \Gamma(\tau^- \to \mu^- (e^-) e^+ e^-),
\ee 
where ${G^{(\prime)}}_{4 \alpha \beta}^{ij} 
= \left( G_\alpha G_\beta/m_{Z^{(\prime)}}^2 \right) \epsilon^{(\prime)}
V_{3i}^* V_{3j}$
with $i,j=1,2,3$ and $\alpha,\beta = L,R$.
Note that ${{G^{(\prime)}}^{ij}_{4RR}}$ and ${{G'}^{ij}_{4LR}}$
are ${\cal O}(\lambda^2)$.
The LFV muon decay rates are described by the same formula
except for the replacement of $V_{32(31)}^* V_{33}$ 
by $V_{31}^* V_{32}$ neglecting electron and muon masses. 

The $\tau^- \to \mu^+ e^- e^-$ and $\tau^- \to e^+ \mu^- \mu^-$ processes 
include only $Z'$-exchange contribution
in the leading order of $\lambda$,
of which decay rates are given by
\be
\Gamma(\tau^- \to \mu^+ e^- e^- (e^+ \mu^- \mu^-)) 
= \frac{m_\tau^5}{96 \pi^3} \left| H_{LL}^{1(2)} \right|^2,
\ee 
where $ {H^{k}_{LL}} = (G'_L/m_{Z'})^2 {\epsilon'}^2 
V_{32}^* V_{33} V_{31}^* V_{3k}$.

Here we discuss a few limiting cases to get lessons on values of $|V_{3i}|$.
For simplicity, the complex phases of the unitary matrix are ignored
in this work.
First, we assume $|V_{33}| \approx 1$. 
Then $|V_{31}|$ and $|V_{32}|$ should be very small
to satisfy the unitarity condition $|V_{31}|^2+ |V_{32}|^2+ |V_{33}|^2=1$
and the strong bounds of the LFV $\tau$ decays,
${\rm Br}(\tau \to l_i l_j l_k) \propto |V_{33}|^2 |V_{3i}|^2
= {\cal O}(10^{-8})$, $i,j,k=1,2$. 
Second, we assume that $|V_{33}| \approx 0$.
Then the LFV $\tau$ decays are automatically suppressed.
We have to satisfy the strong limit on $|V_{31}||V_{32}|$
from the LFV muon decays 
${\rm Br}(\mu \to e e e ) \propto |V_{31}|^2 |V_{32}|^2 < 10^{-12}$
and the unitarity condition $|V_{31}|^2 + |V_{32}|^2 = 1$ 
simulataneously.
The only possible solution is that 
one of $|V_{31}|$ and $|V_{32}|$ should be close to 1
and the other is very small.
Finally, let's consider the case of $|V_{33}|={\cal O}(0.1)$.
Then either $|V_{31}|$ or $|V_{32}|$ 
should also be ${\cal O}(0.1)$ to satisfy the unitarity condition.
In that case, we cannot avoid that
some of the LFV $\tau$ decays involving $|V_{33}| |V_{3i}|$ 
exceed the present experimental bounds.
Therefore we conclude that 
one of the matrix elements $|V_{31}|$, $|V_{32}|$, and $|V_{33}|$ 
is close to 1 and the others are sufficiently small.

When $|V_{3i}|\approx 1$ for an $i$,
the corresponding  $\Gamma(Z \to l_i^- l_i^+)$ data
provide the strong limit on $\epsilon$, and therefore
on $\lambda$ and $\sin^2 \phi$.
For instance, we find that 
$|V_{33}| \approx 1$ leads to the conservative bound $m_{Z'} > 1.8 $ TeV
in Fig. 1.

Now we scan all the parameters $|V_{31}|$, $|V_{32}|$, $|V_{33}|$,
$\sin^2 \phi$ and $m_{Z'}$ 
with the constraints from the CKM unitarity,
the experimental bounds of Eq. (1)-(3), 
and the LEP data of the partial $Z$ widths.
Allowed parameter sets $(\sin^2 \phi, m_{Z'})$
are depicted in Fig. 2.
We find that stronger constraints on the parameters
are obtained from the absence of the LFV processes 
and the partial decay widths of $Z \to l_i^- l_i^+$.
We do not show the bounds from the electroweak precision data
and the low-energy neutral current interaction data in Fig. 2
since their bounds are weaker than that of the CKM unitarity. 
We obtain the conservative lower bound, $m_{Z'}>1.8$ TeV
independent of $\sin^2 \phi$ and $V_{3i}$. 

In conclusion,
we have studied the lepton flavour violating processes in the
extended model based on SU(2)$_l \times$SU(2)$_h \times$U(1)
gauge symmetry.
In this model the FCNC involve the matrix elements, $V_{3i}$, 
as additional parameters,
which diagonalize the charged lepton mass matrix. 
We show that one of $V_{3i}$'s should be close to 1 
while the other two elements are sufficiently small
to accommodate all experimental bounds of the LFV processes
and unitarity condition.
When $V_{3i} \approx 1$, the corresponding partial decay width
of $Z$ boson $\Gamma(Z \to l_i^- l_i^+)$ provides 
strong constraints on the model parameters.
As a result,
we obtain the stronger constraints on this model
from the present experimental data on the LFV and $Z \to l^- l^+$ decays
than those from the previous data.

\acknowledgments
This work was supported 
by WCU program through the KOSEF funded by the MEST (R31-2008-000-10057-0) 
and by the Korea Research Foundation Grant funded 
by the Korean Government (KRF-2008-313-C00167).

\def\PRD #1 #2 #3 {Phys. Rev. D {\bf#1},\ #2 (#3)}
\def\PRL #1 #2 #3 {Phys. Rev. Lett. {\bf#1},\ #2 (#3)}
\def\PLB #1 #2 #3 {Phys. Lett. B {\bf#1},\ #2 (#3)}
\def\NPB #1 #2 #3 {Nucl. Phys. B {\bf #1},\ #2 (#3)}
\def\ZPC #1 #2 #3 {Z. Phys. C {\bf#1},\ #2 (#3)}
\def\EPJ #1 #2 #3 {Euro. Phys. J. C {\bf#1},\ #2 (#3)}
\def\JHEP #1 #2 #3 {JHEP {\bf#1},\ #2 (#3)}
\def\IJMP #1 #2 #3 {Int. J. Mod. Phys. A {\bf#1},\ #2 (#3)}
\def\MPL #1 #2 #3 {Mod. Phys. Lett. A {\bf#1},\ #2 (#3)}
\def\JPG #1 #2 #3 {J. Phys. G {\bf#1},\ #2 (#3)}
\def\PTP #1 #2 #3 {Prog. Theor. Phys. {\bf#1},\ #2 (#3)}
\def\PR #1 #2 #3 {Phys. Rep. {\bf#1},\ #2 (#3)}
\def\RMP #1 #2 #3 {Rev. Mod. Phys. {\bf#1},\ #2 (#3)}
\def\PRold #1 #2 #3 {Phys. Rev. {\bf#1},\ #2 (#3)}
\def\IBID #1 #2 #3 {{\it ibid.} {\bf#1},\ #2 (#3)}

\end{document}